\documentclass[english,prb,floatfix,twocolumn,showpacs]{revtex4}
\usepackage[T1]{fontenc}
\usepackage[latin9]{inputenc}
\usepackage{float}
\usepackage{amsmath}
\usepackage{amssymb}
\usepackage{graphicx}
\usepackage{esint}

\makeatletter
\@ifundefined{textcolor}{}
{%
 \definecolor{BLACK}{gray}{0}
 \definecolor{WHITE}{gray}{1}
 \definecolor{RED}{rgb}{1,0,0}
 \definecolor{GREEN}{rgb}{0,1,0}
 \definecolor{BLUE}{rgb}{0,0,1}
 \definecolor{CYAN}{cmyk}{1,0,0,0}
 \definecolor{MAGENTA}{cmyk}{0,1,0,0}
 \definecolor{YELLOW}{cmyk}{0,0,1,0}
}

\makeatother

\usepackage{babel}
\begin{document}

\title{Interplay of classical and ``quantum'' capacitance in a one dimensional
array of Josephson junctions}

\author{Pedro Ribeiro}

\affiliation{CFIF, Instituto Superior Técnico, Universidade de Lisboa, Av. Rovisco
Pais, 1049-001 Lisboa, Portugal}

\affiliation{Max Planck Institute for the Physics of Complex Systems - Nöthnitzer
Str. 38, , D-01187 Dresden, Germany}

\affiliation{Max Planck Institute for Chemical Physics of Solids - Nöthnitzer
Str. 40, D-01187 Dresden, Germany}

\author{Antonio M. García-García}

\affiliation{University of Cambridge, Cavendish Laboratory, JJ Thomson Avenue,
Cambridge, CB3 0HE, UK }

\affiliation{CFIF, Instituto Superior Técnico, Universidade de Lisboa, Av. Rovisco
Pais, 1049-001 Lisboa, Portugal}
\begin{abstract}
Even in the absence of Coulomb interactions phase fluctuations induced
by quantum size effects become increasingly important in superconducting
nano-structures as the mean level spacing becomes comparable with
the bulk superconducting gap. Here we study the role of these fluctuations,
termed ``quantum capacitance'', in the phase diagram of a one-dimensional
(1D) ring of ultrasmall Josephson junctions (JJ) at zero temperature
by using path integral techniques. Our analysis also includes dissipation
due to quasiparticle tunneling and Coulomb interactions through a
finite mutual and self capacitance. The resulting phase diagram has
several interesting features: A finite quantum capacitance can stabilize
superconductivity even in the limit of only a finite mutual-capacitance
energy which classically leads to breaking of phase coherence. In
the case of vanishing charging effects, relevant in cold atom settings
where Coulomb interactions are absent, we show analytically that superfluidity
is robust to small quantum finite-size fluctuations and identify the
minimum grain size for phase coherence to exist in the array. We have
also found that the renormalization group results are in some cases
very sensitive to relatively small changes of the instanton fugacity.
For instance, a certain combination of capacitances could lead to
a non-monotonic dependence of the superconductor-insulator transition
on the Josephson coupling.
\end{abstract}

\pacs{74.20.Fg, 75.10.Jm, 71.10.Li, 73.21.La}

\maketitle
\global\long\def\ket#1{\left| #1\right\rangle }

\global\long\def\bra#1{\left\langle #1 \right|}

\global\long\def\kket#1{\left\Vert #1\right\rangle }

\global\long\def\bbra#1{\left\langle #1\right\Vert }

\global\long\def\braket#1#2{\left\langle #1\right. \left| #2 \right\rangle }

\global\long\def\bbrakket#1#2{\left\langle #1\right. \left\Vert #2\right\rangle }

\global\long\def\av#1{\left\langle #1 \right\rangle }

\global\long\def\tr{\text{Tr}}

\global\long\def\pd{\partial}

\global\long\def\im{\text{Im}}

\global\long\def\re{\text{Re}}

\global\long\def\sgn{\text{sgn}}

\global\long\def\Det{\text{Det}}

\global\long\def\abs#1{\left|#1\right|}

\global\long\def\up{\uparrow}

\global\long\def\down{\downarrow}

\global\long\def\grad{\nabla}

\global\long\def\bs#1{\boldsymbol{#1}}

The Josephson's effect, \cite{jos,giaver} reveals the central role
played by the phase of the order parameter in superconductivity. It
has been exploited in a broad spectrum of research problems and applications:
from the study of the pseudogap phase in high $T_{c}$ materials \cite{pseuhtc}
, fluctuations above $T_{c}$ \cite{scalapino} and cold atom physics
\cite{cold} to spintronics \cite{spint} and quantum computing \cite{qinf}.
Of special interest is the study of an array of superconducting grains
separated by thin tunnel junctions, usually referred to as Josephson
junctions (JJ). The physical properties of JJ arrays are very sensitive
to the grain dimensionality, the presence of Coulomb interactions
and dissipation \cite{fisher,kivelson,shoen2d,doniach,efetov} (see
also the review \cite{revjj}). Usually it is assumed that each single
grain is sufficiently large so that the amplitude of the order parameter,
the superconducting gap, is well described by the bulk Bardeen-Cooper-Schriffer
(BCS) theory. Moreover it is also commonly assumed that a simple capacitance
model is sufficient to account for Coulomb interactions. The phase
of each grain is therefore the only effective degree of freedom of
the JJ array. 

Within this general theoretical framework a broad consensus has emerged
on the main features of JJ arrays: for long 1d arrays at zero temperature
with negligible dissipation, the existence of long range order depends
on the nature of the capacitance interactions. For situations in which
only self-capacitance is important superconductivity persists for
sufficiently small charging effects \cite{doniach} provided that
the Josephson coupling is strong enough. Despite spatial global long-range
order a state of zero resistance will strictly occur only in the case
in which the super current is induced by threading a flux in a ring-shaped
JJ array \cite{fazio2,zaikin}. A current in a long but finite linear
JJ array will eventually induce a resistance though for sufficiently
strong Josephson coupling it is hard to measure it as its typical
time scale can be much longer that the experimental observation time.
At any finite temperature the resistivity is always finite as a consequence
of the unbinding of phase anti-phase slips. 

In the opposite limit in which only mutual-capacitance is considered,
even small charging effects induce a superconductor insulator transition.
The combined effect of the two types of charging effects, considered
in \cite{newcap}, can also lead to global long-range order. On a
single junction, dissipation by quasiparticle tunneling only renormalizes
\cite{schoen} the value of the capacitance. However dissipation caused
by a ohmic resistance \cite{caldeira} induces long range correlations
between phase slips and anti phase slips that restore superconductivity
provided that the normal resistance is smaller than the quantum one.
In order to illustrate the profound impact of dissipation it is worth
noting that a state of zero resistance in a 1D JJ array can in some
cases coexist \cite{fisher} with an order parameter whose spatial
correlation functions are short-ranged. 

The closely related problem of a quantum nanowire was addressed in
\cite{zaikin,blatter} by employing instanton techniques to model
phase tunneling and then mapping the resulting effective model onto
a 1+1d Coulomb gas where one of the dimensions is imaginary time.
For an infinite wire in the zero temperature limit a superconductor-insulator
Berezinsky-Kosterlitz-Thouless (BKT) transition occurs as a function
of the system parameters. The role of vortices in 1+1d is played by
phase slips which correspond to configurations for which the amplitude
of the order parameter vanishes and the phase receives a $2\pi$ boost.
By contrast at finite temperature -- a similar argument holds for
finite length -- the time dimension is compactified so, in the absence
of dissipation, the Coulomb gas analogy breaks down since for long
separations phase and anti-phase slips become uncorrelated. As a consequence
phase coherence is lost and the resistance is always finite \cite{halperin,zaikin,newcap}. 

As was mentioned previously all these results assume that the amplitude
of the order parameter of each grain, which enters in the definition
of the Josephson coupling energy, is not affected by any deviations
from the bulk limit and that the phase dynamics is induced only by
classical charging effects. Although these assumptions are in many
cases sound there are situations in which corrections are expected. 

In sufficiently small grains close to the critical temperature it
is well documented that homogeneous path integral configurations different
from the mean field prediction, the so called static paths, contribute
significantly to the specific heat and other thermodynamical observables
\cite{scalapino1}. For single nano-grains at intermediate temperatures
it has been shown recently \cite{us} that, even in the limit of vanishing
Coulomb interactions, deviations from mean-field predictions occur
due to the non trivial interplay of thermal and quantum fluctuations
induced by finite size effects. Experimentally it is also well established
\cite{tinkham,nmat} that substantial deviations from mean-field predictions
occur in isolated nano-grains. Indeed it has recently been reported
\cite{nmat,altprl} that quantum size effects enhance the superconducting
gap of single isolated Sn nanograins with respect to the bulk limit. 

It is therefore of interest to understand in more detail the role
of these finite size effects in arrays of ultrasmall JJ where the
mean level energy spacing of single grains is smaller, but comparable,
to the superconducting gap. This paper is a step in this direction.
We study the stability of phase coherence in arrays of 1D JJ at zero
temperature. Our formalism includes the above quantum fluctuations
induced by size effects, charging effects and dissipation by quasiparticle
tunneling. Starting from a microscopic Hamiltonian for a 1D JJ ring-shaped
array of nanograins at zero temperature, we map the problem onto a
Sine Gordon Hamiltonian where we identify the region of parameters
in which long-range order persists in the presence of phase fluctuations.
In the limit of vanishing charging energy, relevant for cold atom
experiments, we find the minimum size for which the JJ array can be
superfluid as a function of the wire resistance in the normal state.
We also show that quantum fluctuations induced by finite size effects
can in principle stabilize superconductivity in the limit of a negligible
self-capacitance energy but a finite mutual capacitance energy. We
have also identified a region parameters in which it is observed a
non-monotonic dependence of the superconductor-insulator transition
on the Josephson coupling.

\section{The model}

\begin{figure}
\centering{}\includegraphics[width=1\columnwidth]{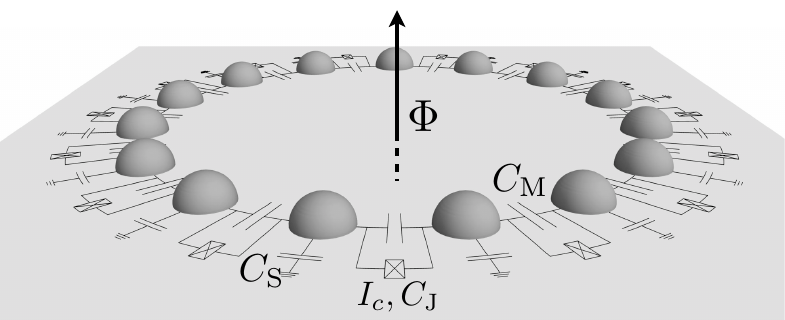}\caption{\label{fig:system}Sketch a closed ring of Josephson junctions pierced
by a total flux $\Phi$. }
\end{figure}

We consider the system sketched in Fig.(\ref{fig:system}), consisting
of an array of $L$ superconducting grains with periodic boundary
conditions and a total magnetic flux $\Phi$ passing through it, that
can be modeled by the Hamiltonian: 
\begin{align}
H= & \sum_{r=1}^{L}H_{r}^{\text{BCS}}+H_{r}^{\text{SC}}+H_{r,r+1}^{\text{MC}}+H_{r,r+1}^{\text{T}}.\label{eq:H_Junction}
\end{align}
Each isolated superconducting grain is described by the BCS term, 

\begin{alignat}{1}
H_{r}^{BCS} & =\sum_{\alpha,\sigma}\epsilon_{\alpha,r}\, c_{\alpha,\sigma,r}^{\dagger}c_{\alpha,\sigma,r}\label{eq:Hamilton}\\
 & -g_{r}\delta_{r}\left(\sum_{\alpha}c_{\alpha,1,r}^{\dagger}c_{-\alpha,-1,r}^{\dagger}\right)\left(\sum_{\alpha'}c_{-\alpha',-1,r}c_{\alpha',1,r}\right),\nonumber 
\end{alignat}
accounting for the effective attractive electron-electron interactions
in the region where the grain size is much smaller then the bulk superconducting
coherence length. $\alpha,-\alpha$ label single-particle states related
by time reversal symmetry with energies $\epsilon_{\alpha}=\epsilon_{-\alpha}$,
$\sigma=\pm1$ is the spin label and $\delta_{r}$ and $g_{r}$ are,
respectively, the mean level spacing (inversely proportional to the
grain volume) and the dimensionless coupling constant of grain $r$.
We further assume the presence of self and mutual capacitive terms
of the form
\begin{align}
H_{r}^{\text{S}} & =\frac{1}{2C_{r}^{\text{S}}}\left(\hat{N}_{r}-N_{r}^{\text{S}}\right)^{2},\\
H_{r,r+1}^{\text{M}} & =\frac{1}{2C_{r}^{M}}\left(\hat{N}_{r}-\hat{N}_{r+1}-N_{r}^{\text{M}}\right)^{2},
\end{align}
accounting for the repulsive Coulomb interaction within each grain
and between electrons in neighboring grains. $\hat{N}_{r}=\sum_{\alpha,\sigma}c_{\alpha,\sigma,r}^{\dagger}c_{\alpha,\sigma,r}$
is the total number of electrons, $C_{r}^{\text{S}}$ is the self-capacitance
the of grain $r$ and $C_{r}^{M}$ the mutual capacitance between
nearest neighbor grains $r$ and $r+1$. The constants $N_{r}^{\text{S}}$
and $N_{r}^{\text{M}}$ can be adjusted by applying suitable gate
voltages. Finally, the hopping of electrons between grains is captured
by the term 
\begin{multline}
H_{r,r+1}^{\text{T}}=\sum_{\alpha\alpha'\sigma}T_{r,r+1}^{\alpha,\alpha'}c_{\alpha,\sigma,t}^{\dagger}c_{\alpha',\sigma,r+1}+\text{h.c.},
\end{multline}
where the hybridization matrix $T_{r,r+1}^{\alpha,\alpha'}\propto\int\psi_{\alpha,\sigma,r}\left(\bs x\right)\bar{\psi}_{\alpha',\sigma,r+1}\left(\bs x\right)d\bs x$
is proportional to the overlap of the single-particle wave functions
of two neighboring grains. In the regime of interest here - small
grain sizes with respect to the bulk coherence length - the simplifying
assumption that the hybridization is energy independent $T_{r,r+1}^{\alpha,\alpha'}=t_{r,r+1}$
can safely be used and thus $H_{r,r+1}^{\text{T}}$ simplifies to
\begin{multline}
H_{r,r+1}^{\text{T}}=t_{r,r+1}\sum_{\sigma}\left(\sum_{\alpha}c_{\alpha,\sigma,r}^{\dagger}\right)\left(\sum_{\alpha}c_{\alpha,\sigma,r+1}\right)+\text{h.c.},
\end{multline}
with $\Phi=\sum_{i}\arg t_{r,r+1}$ the total flux passing through
the ring.

\section{Finite size corrections to the action of a Josephson Junction's Array}

\subsection{Partition function in the path integral formalism}

In this section we write the partition function $Z=\tr\left[e^{-\beta H}\right]$
in the path-integral form and identify the finite size corrections
to the action. This is done by inserting $L$ complex-valued Hubbard-Stratonovich
fields (HSF) $\Delta_{r}$ to decouple the BCS term in the superconducting
channel, $L$ real valued HSF $V_{r}^{\text{S}}$, conjugate to the
number of particles on each grain, and $L$ real valued HSF $V_{r}^{\text{M}}$,
conjugate to the difference of the number of particles in neighboring
grains. Using the notation $\Psi=\left(c_{\alpha,1,1},c_{\alpha,-1,1}^{\dagger},c_{\alpha,1,2},c_{\alpha,-1,2}^{\dagger},...\right)^{T}$,
the partition function reads $Z=\int Dc\, D\Delta\, DV\, e^{-S}$,
with the action 
\begin{multline}
S=-\Psi^{\dagger}G^{-1}\Psi+\int_{0}^{\beta}d\tau\sum_{r}\left[\frac{1}{g_{r}\delta_{r}}\Delta_{r}^{\dagger}\Delta_{r}\right.\\
\left.+\frac{C_{r}^{\text{S}}}{2}\left(V_{r}^{\text{S}}\right)^{2}+iN_{r}^{\text{S}}V_{r}^{\text{S}}+\frac{C_{r}^{\text{M}}}{2}\left(V_{r}^{\text{M}}\right)^{2}+iN_{r}^{\text{M}}V_{r}^{\text{M}}\right],
\end{multline}
where the full Green's function is given by
\begin{eqnarray}
G^{-1} & = & \left(\begin{array}{ccc}
G_{1}^{-1} & T_{21}\\
T_{21}^{\dagger} & G_{2}^{-1} & \ddots\\
 & \ddots & \ddots
\end{array}\right),
\end{eqnarray}
and 
\begin{eqnarray}
G_{r}^{-1} & = & \left(\begin{array}{cc}
-\partial_{\tau}-\tilde{\varepsilon}_{\alpha,r}\left(\tau\right) & \Delta_{r}\left(\tau\right)\\
\Delta_{r}^{\dagger}\left(\tau\right) & -\partial_{\tau}+\tilde{\varepsilon}_{\alpha,r}\left(\tau\right)
\end{array}\right),
\end{eqnarray}
is the inverse of the electronic propagators restricted to grain $r$.
Here we defined $\tilde{\varepsilon}_{\alpha,r}\left(\tau\right)=\varepsilon_{\alpha,r}-iV_{r}^{\text{S}}\left(\tau\right)-iV_{r}^{\text{M}}\left(\tau\right)+iV_{r-1}^{\text{M}}\left(\tau\right)$
and the hybridization matrix $T_{r+1,r}=\left(\begin{array}{cc}
t_{r+1,r} & 0\\
0 & -\bar{t}_{r+1,r}
\end{array}\right)$. 

Integrating out $\Psi$ yields the action 
\begin{multline}
S=-\tr\ln\left[-G^{-1}\right]+\int_{0}^{\beta}d\tau\sum_{r}\left[\frac{1}{g_{i}\delta_{r}}\Delta_{r}^{\dagger}\Delta_{r}\right.\\
\left.+\frac{C_{r}^{\text{S}}}{2}\left(V_{r}^{\text{S}}\right)^{2}+iN_{r}^{\text{S}}V_{r}^{\text{S}}+\frac{C_{r}^{\text{M}}}{2}\left(V_{r}^{\text{M}}\right)^{2}+iN_{r}^{\text{M}}V_{r}^{\text{M}}\right]
\end{multline}
solely in terms of the HSF. 

We apply the unitary transformation
\begin{align*}
U & =\text{diag}\left\{ e^{i\frac{1}{2}\phi_{1}\left(\tau\right)},e^{-i\frac{1}{2}\phi_{1}\left(\tau\right)},e^{i\frac{1}{2}\phi_{2}\left(\tau\right)},e^{-i\frac{1}{2}\phi_{2}\left(\tau\right)},\ldots\right\} 
\end{align*}
with $\phi_{r}\left(\tau\right)=\phi_{r}\left(\tau+\beta\right)+2\pi n_{\phi_{r}}$
$\left(n_{\phi_{r}}\in\mathbb{Z}\right)$ to the electronic propagator
$G^{-1}$ in order to render real its off-diagonal anomalous elements
$\Delta_{r}\left(\tau\right)=s_{r}\left(\tau\right)e^{i\phi_{r}\left(\tau\right)}$,
where $s_{r}\left(\tau\right),\phi_{r}\left(\tau\right)\in\mathbb{R}$.
Note that for odd $n_{\phi_{i}}$ one has that $\tr_{f}\left[G^{-1}\right]=\tr_{b}\left[U^{\dagger}G^{-1}U\right]$,
where $\tr_{f}$ denotes the trace over anti-periodic functions (fermionic)
and $\tr_{b}$ the trace over periodic functions (bosonic). For a
generic $n_{\phi_{r}}$ we will denote $\tr_{n_{\phi_{r}}}=\tr_{f}$
for $n_{\phi_{r}}$ even and $\tr_{n_{\phi_{r}}}=\tr_{b}$ for $n_{\phi_{r}}$
odd. Whenever we have two such indices we will use $\tr_{n_{\phi_{1}}n_{\phi_{2}}}$
for the time periodicity in indices $1$ and $2$. Note however that
this complication is only formal as we will be interested in the low
temperature properties of this action where the distinction between
even and odd $n_{\phi}$'s can be safely ignored \cite{matveev}.
After this transformation we get 
\begin{eqnarray}
\tilde{G}^{-1} & = & U^{\dagger}G^{-1}U=\left(\begin{array}{ccc}
\tilde{G}_{1}^{-1} & \tilde{T}_{21}\\
\tilde{T}_{21}^{\dagger} & \tilde{G}_{2}^{-1} & \ddots\\
 & \ddots & \ddots
\end{array}\right),
\end{eqnarray}
with
\begin{multline}
\tilde{G}_{r}^{-1}=-1\times\\
\left(\begin{array}{cc}
\partial_{\tau}+\tilde{\varepsilon}_{\alpha,r}\left(\tau\right)+i\frac{1}{2}\partial_{\tau}\phi_{r}\left(\tau\right) & -s_{r}\left(\tau\right)\\
-s_{r}\left(\tau\right) & \partial_{\tau}-\tilde{\varepsilon}_{\alpha,r}\left(\tau\right)-i\frac{1}{2}\partial_{\tau}\phi_{r}\left(\tau\right)
\end{array}\right),
\end{multline}
and
\begin{multline}
\tilde{T}_{r+1,r}=\\
\left(\begin{array}{cc}
t_{r+1,r}e^{i\frac{1}{2}\left[\phi_{r+1}\left(\tau\right)-\phi_{r}\left(\tau\right)\right]} & 0\\
0 & -\bar{t}_{r+1,r}e^{-i\frac{1}{2}\left[\phi_{r+1}\left(\tau\right)-\phi_{r}\left(\tau\right)\right]}
\end{array}\right).
\end{multline}
 Moreover, assuming the hopping amplitude to be small, we may develop
the $\tr\ln\left[-G^{-1}\right]$ term to second order in $\abs{t_{r+1,r}}$
and obtain the action 
\begin{multline}
S\left[s,\phi,V\right]=\sum_{r}\left\{ \int_{0}^{\beta}d\tau\left[\frac{1}{g_{i}\delta_{r}}s_{r}^{\dagger}s_{r}\right.\right.\\
\left.+\frac{C_{r}^{\text{S}}}{2}\left(V_{r}^{\text{S}}\right)^{2}+iN_{r}^{\text{S}}V_{r}^{\text{S}}+\frac{C_{r}^{\text{M}}}{2}\left(V_{r}^{\text{M}}\right)^{2}+iN_{r}^{\text{M}}V_{r}^{\text{M}}\right]\\
-\tr_{n_{\phi_{r}}}\ln\left[-\tilde{G}_{r}^{-1}\right]\\
\left.+\tr_{n_{\phi_{r}}n_{\phi_{r+1}}}\left[\tilde{G}_{r+1}\tilde{T}_{r+1,i}\tilde{G}_{r}\tilde{T}_{r,r+1}^{\dagger}\right]\right\} .\label{eq:Action_2}
\end{multline}

\subsection{Leading behavior in $\delta$}

The action above Eq.(\ref{eq:Action_2}) is suitable for a saddle-point
expansion in both $s$ and $V$ fields since the action for each grain
is an extensive quantity in the number of electrons within that grain
$\av{N_{r}}\simeq E_{D}/\delta_{r}$ . Notice however that the saddle-point
equations cannot be explicitly evaluated as $\tilde{G}^{-1}$ depends
on $\phi_{r}\left(\tau\right)$. We proceed by noting that $\partial_{\tau}\phi_{r}\left(\tau\right)$
is small, as the phase varies smoothly as a function of $\tau$ for
sufficiently low temperatures. Formally we set $V_{r}^{\text{S}}\left(\tau\right)=V_{r,0}^{\text{S}}+\delta V_{r}^{\text{S}}\left(\tau\right)$,
$V_{r}^{M}\left(\tau\right)=V_{r,0}^{M}+\delta V_{r}^{M}\left(\tau\right)$
and $s_{r}\left(\tau\right)=s_{r,0}+\delta s_{r}\left(\tau\right)$
where the subscript $0$ denotes the static component (constant in
$\tau$) of the different quantities and the fluctuation around the
static value, to be considered at quadratic order, are denoted by
$\delta V_{r}^{\text{S}}$, $\delta V_{r}^{\text{M}}$ and $\delta s_{r}$.
Physically, $s_{r,0}$ is the amplitude of the condensate on grain
$i$ and the terms $iV_{r,0}^{\text{S}},iV_{r,0}^{\text{M}}\in\mathbb{R}$
leads to a renormalization of the chemical potential: $\tilde{\varepsilon}_{\alpha,r}=\varepsilon_{\alpha,r}-iV_{r,0}^{\text{S}}-iV_{r,0}^{\text{M}}+iV_{r-1,0}^{\text{M}}$
. 

For equally spaced levels and a particle-hole-symmetric single-particle
density of states the tunneling term can be simplified at low temperatures
\cite{ambegaokar} 
\begin{multline*}
\tr\left[\tilde{G}_{r+1}\tilde{T}_{r+1,r}\tilde{G}_{r}\tilde{T}_{r,r+1}^{\dagger}\right]\simeq\\
\frac{C_{i}^{\text{J}}}{8}\int d\tau\left\{ \pd_{\tau}\left[\phi_{r+1}\left(\tau\right)-\phi_{r}\left(\tau\right)\right]\right\} ^{2}\\
-\frac{I_{i}^{\text{c}}}{2}\int d\tau\cos\left[\phi_{r+1}\left(\tau\right)-\phi_{r}\left(\tau\right)+\phi_{r}^{t}\right]
\end{multline*}
where $\phi_{r}^{t}$ is the phase of the hopping term $t_{r+1,r}=\abs{t_{r+1,r}}e^{i\phi_{r}^{t}}$,
$C_{r}^{\text{J}}$ is the quasi-particle induced capacitance and
$I_{r}^{\text{c}}$ is the junction's critical current between grains
$r$ and $r+1$, given respectively by \cite{eckern}
\begin{multline}
C_{i}^{\text{J}}=2\frac{4\abs{t_{r+1,r}}^{2}}{\delta_{r}\delta_{r+1}}\times\\
\int_{s_{r,0}}^{\infty}d\nu_{1}\int_{s_{r+1,0}}^{\infty}d\nu_{2}\frac{\nu_{1}\nu_{2}}{\left(\nu_{2}+\nu_{1}\right)^{3}\sqrt{\left(\nu_{1}^{2}-s_{r,0}^{2}\right)\left(\nu_{1}^{2}-s_{r+1,0}^{2}\right)}}
\end{multline}
and
\begin{multline}
I_{r}^{\text{c}}=\frac{8\abs{t_{r+1,r}}^{2}}{\delta_{r}\delta_{r+1}}\times\\
\int_{s_{r,0}}^{\infty}d\nu_{1}\int_{s_{r+1,0}}^{\infty}d\nu_{2}\frac{s_{r,0}s_{r+1,0}}{\left(\nu_{2}+\nu_{1}\right)\sqrt{\left(\nu_{1}^{2}-s_{r,0}^{2}\right)\left(\nu_{1}^{2}-s_{r+1,0}^{2}\right)}}
\end{multline}
Note that for $s_{r,0}=s_{r+1,0}=s_{0}$ these expressions simplify
to $C_{r}^{\text{J}}=C_{\text{J}}=\frac{3\pi}{32}\frac{1}{s_{0}R_{N}}$
and $I_{r}^{\text{c}}=I_{\text{c}}=\frac{\pi}{2}\frac{s_{0}}{R_{N}}$
with $R_{N}=\left(\frac{4\abs t^{2}\pi}{\delta^{2}}\right)^{-1}$
the normal state resistance of the junction.

With these approximations the action reads 
\begin{multline}
S\left[s,\phi,V\right]=S_{0}+\int d\tau\sum_{r}\left\{ \Omega_{r}\delta s_{r}^{2}\left(\tau\right)\right.\\
+\frac{C_{r}^{\text{S}}}{2}\delta V_{r}^{\text{S}}\left(\tau\right)^{2}+\frac{C_{r}^{\text{M}}}{2}\delta V_{r}^{\text{M}}\left(\tau\right)^{2}+\frac{1}{2}C_{\delta,r}\varphi_{r}^{2}\left(\tau\right)\\
-i\av{N_{0,r}}\pd_{\tau}\phi_{r}\left(\tau\right)+\frac{C_{r}^{\text{J}}}{8}\left(\pd_{\tau}\left[\phi_{r+1}\left(\tau\right)-\phi_{r}\left(\tau\right)\right]\right)^{2}\\
\left.-\frac{I_{r}^{\text{c}}}{2}\cos\left[\phi_{r+1}\left(\tau\right)-\phi_{r}\left(\tau\right)+\phi_{r}^{t}\right]\right\} \label{eq:Action_3-1}
\end{multline}
where
\begin{multline}
S_{0}=\sum_{i}\left\{ \tr\ln\left[-\tilde{G}_{r,0}^{-1}\right]\right.\\
+\beta\left[\frac{1}{g_{r}\delta_{r}}s_{r,0}^{2}+\frac{C_{r}^{\text{S}}}{2}\left(V_{r,0}^{\text{S}}\right)^{2}+iN_{r}^{\text{S}}V_{r,0}^{\text{S}}\right.\\
\left.\left.+\frac{C_{r}^{\text{M}}}{2}\left(V_{r,0}^{\text{M}}\right)^{2}+iN_{r}^{\text{M}}V_{r,0}^{\text{M}}\right]\right\} 
\end{multline}
only depend on the static saddle-point values, $\Omega_{r}=\frac{1}{g_{r}\delta_{r}}-\frac{1}{2}\left(\sum_{\alpha}\frac{\tilde{\varepsilon}_{\alpha,r}^{2}}{\xi_{\alpha,r}^{3}}\right)$
, $\xi_{\alpha,r}=\sqrt{\tilde{\varepsilon}_{\alpha,r}^{2}+s_{0,r}^{2}}$
, $\av{N_{0,r}}=\frac{1}{2}\sum_{\alpha}\left(1-\frac{\tilde{\varepsilon}_{\alpha,r}}{\xi_{\alpha,r}}\right)$,
\begin{eqnarray}
\tilde{G}_{r,0}^{-1}\left(i\omega_{n}\right) & = & \left(\begin{array}{cc}
i\omega_{n}-\tilde{\varepsilon}_{\alpha,r} & s_{r,0}\\
s_{r,0} & i\omega_{n}+\tilde{\varepsilon}_{\alpha,r}
\end{array}\right)
\end{eqnarray}
and where we also define
\begin{multline}
\varphi_{r}\left(\tau\right)=\delta V_{r}^{\text{S}}\left(\tau\right)+\delta V_{r}^{\text{M}}\left(\tau\right)-\delta V_{r-1}^{\text{M}}\left(\tau\right)-\frac{1}{2}\partial_{\tau}\phi_{r}\left(\tau\right)
\end{multline}

and the finite size induced self-capacitance $C_{\delta,r}=\frac{2}{\delta_{r}}$. 

Eq. (\ref{eq:Action_3-1}) is now suitable to a static-path treatment
\cite{us} once the fluctuations are integrated out. Here, as we are
only interested in the phase dynamics at low temperatures we set the
static components to their mean-field values and integrate out the
gapped fluctuations both in the $s$ and $V$ fields. In the limit
$\beta s_{0}\gg1$ the final action in terms of the phase degrees
of freedom and assuming translational invariance in the couplings
$C_{r}^{\text{S}}=C_{\text{S}}$, $C_{r}^{\text{M}}=C_{\text{M}}$,
$C_{\delta,r}=C_{\delta}$, is given by, 
\begin{alignat}{1}
S & =\nonumber \\
 & \frac{1}{8}\int d\tau\sum_{r,r'}\partial_{\tau}\phi_{r}\left(\tau\right)\left[C_{R}\right]_{r,r'}\partial_{\tau}\phi_{r'}\left(\tau\right)\nonumber \\
 & +i\frac{1}{2}\sum_{r}\av{N_{0,r}}\int d\tau\pd_{\tau}\phi_{r}\left(\tau\right)\nonumber \\
 & -\frac{I^{\text{c}}}{2}\sum_{r}\int d\tau\cos\left[\phi_{r+1}\left(\tau\right)-\phi_{r}\left(\tau\right)+\phi_{r}^{t}\right]\label{eq:Action_4}
\end{alignat}
where $C_{R}=\frac{1}{\tilde{C}_{\text{S}}^{-1}-C_{\text{M}}^{-1}\Delta_{1}^{2}}-C_{\text{J}}\Delta_{1}^{2}$
is the capacitance matrix, with $\Delta_{1}$ the discrete derivative:
$\left(\Delta_{1}\phi\right)_{r}=\phi_{r}-\phi_{r-1}$, $\phi_{r}^{t}=\arg t_{r,r+1}$
is the phase of the hopping term, $\av{N_{0,r}}=\frac{1}{2}\sum_{\alpha}\left(1-\frac{\tilde{\varepsilon}_{\alpha,r}}{\xi_{\alpha,r}}\right)$
is the average number of electrons in grain $r$ and 
\begin{eqnarray}
\tilde{C}_{\text{S}} & = & \left(\frac{1}{C_{\text{S}}}+\frac{\delta}{2}\right)^{-1}
\end{eqnarray}
 is the grain self-capacitance renormalized by quantum finite size
effects. Note that on the lattice $\sum_{r}\left(\Delta_{1}\phi\right)_{r}\left(\Delta_{1}\phi'\right)_{r}=-\sum_{r}\phi_{r}\left(\bar{\Delta}_{1}\Delta_{1}\phi'\right)_{j}$,
with $\left(\bar{\Delta}_{1}\phi\right)_{r}=\phi_{r+1}-\phi_{r}$,
for sake of simplicity we use the notation $\Delta_{1}^{2}$ to denote
the lattice Laplacian $\bar{\Delta}_{1}\Delta_{1}$. 

Eq.(\ref{eq:Action_4}) is the central result of this section, it
contains the effective low energy theory for a junction at $T\ll T_{c}$,
including charging effects, quasiparticle dissipation and for the
first time quantum fluctuations induced by finite size effects $C_{\delta}$.
The Berry phase term - second term of Eq.(\ref{eq:Action_4}) - ensures
that, in the ground-state (i.e. for $T=0$), the average number of
electrons on each grain is even \cite{matveev}. In the following
we assume that this condition is fulfilled and drop this term. 

Note that for a set of isolated finite-size grains with $I^{\text{c}}=C_{\text{J}}=C_{\text{M}}^{-1}=0$
no superconducting phase ensues as the action in Eq.(\ref{eq:Action_4})
reduces to $\frac{\varrho}{2}\int d\tau\left[\pd_{\tau}\phi_{r}\left(\tau\right)\right]^{2}$
with the phase stiffness $\varrho=\frac{\tilde{C}_{\text{S}}}{4}$
controlling the exponential time decay of the order parameter correlation
function $\hat{\Psi}_{r}\left(\tau\right)=g\delta\sum_{\alpha}c_{r,\alpha}\left(\tau\right)c_{r,\bar{\alpha}}\left(\tau\right)$:
$\av{\Psi_{r}\left(\tau\right)\Psi_{r'}^{\dagger}\left(\tau'\right)}\propto\delta_{r,r'}\, s_{0}^{2}e^{-\frac{\abs{\tau-\tau'}}{2\varrho}}$.

\section{Superconducting Transition}

\subsection{Hamiltonian Formulation}

In this section we analyze the action given by Eq.(\ref{eq:Action_4}),
without the Berry phase term $\int d\tau\pd_{\tau}\phi_{r}\left(\tau\right)$
as we assume an even number of electrons in each grain. The calculation
is carried out by first mapping this equation onto an equivalent Coulomb
gas model. The Coulomb gas is subsequently transformed into a Sine-Gordon
action for which a perturbative RG treatment can be effectively performed. 

First we provide a description of the model in terms of the effective
low energy Hamiltonian for the phase degrees of freedom in order to
make contact with previous works where this effective description
is taken as the starting point of the calculation. The initial step
is the discretization of the imaginary time in Eq.(\ref{eq:Action_4}):
$\tau=\Delta\tau\,\tilde{\tau}$ (with $\tilde{\tau}=1,...,N$ and
$N\Delta\tau=\beta$) . Using the identity 
\begin{multline}
\lim_{\Delta\tau\to0}\sum_{\bs n=n_{1},..,n_{N}}e^{-\frac{\Delta\tau}{2}\bs n.\bs A^{-1}.\bs n+\Delta\tau i\bs b\bs n}=\\
\left(\sqrt{\frac{2\pi}{\det\left(\bs A\right)\Delta\tau}}\right)^{N}e^{-\frac{\Delta\tau}{2}\bs b.\bs A.\bs b}
\end{multline}
the partition function can be rewritten as $Z=\int D\phi\sum_{n}e^{-iS\left[\phi,n\right]}$,
with 
\begin{alignat}{1}
S\left[\phi,n\right] & =\sum_{\tilde{\tau},r,r'}2\Delta\tau\, n\left(\tilde{\tau},r\right)\left[C_{R}^{-1}\right]_{rr'}n\left(\tilde{\tau},r'\right)\nonumber \\
 & -\sum_{\tilde{\tau},r}in\left(\tilde{\tau},r\right)\left[\phi\left(\tilde{\tau}+1,r\right)-\phi\left(\tilde{\tau},r\right)\right]\nonumber \\
 & -\frac{I_{c}}{2}\sum_{\tilde{\tau},r}\Delta\tau\cos\left[\phi\left(\tilde{\tau},r+1\right)-\phi\left(\tilde{\tau},r\right)+\phi_{r}^{t}\right]\label{eq:Action_JJ_2}
\end{alignat}
In this form, Eq.(\ref{eq:Action_JJ_2}) can readily be interpreted
as the Trotter-sliced action coming from the Hamiltonian 
\begin{eqnarray}
H & = & \sum_{rr'}2\hat{n}_{r}\left[C_{R}^{-1}\right]_{rr'}\hat{n}_{r'}\nonumber \\
 &  & -\frac{I_{c}}{2}\sum_{r}\cos\left[\hat{\phi}_{r+1}-\hat{\phi}_{r}+\phi_{r}^{t}\right]
\end{eqnarray}
where $\hat{n}_{r}=\left(-i\pd_{\phi_{r}}\right)$, the variable conjugated
to $\hat{\phi}_{r}$, is the number of Cooper-pairs in grain $r$.

\subsection{Partition function of the Coulomb gas \label{sub:Partition-function-of}}

We follow the procedure of \cite{jose} to re-write the action of
a Josephson junction array in terms of the partition function of a
classical Coulomb gas. Using the Villain decomposition of the cosine
term 
\begin{eqnarray}
e^{z\cos\left(\theta\right)} & \simeq & I_{0}\left(z\right)\sum_{m=-\infty}^{\infty}e^{-\frac{1}{2}\mu\left(z\right)m^{2}}e^{im\theta}\label{eq:Villain}
\end{eqnarray}
with $I_{0}\left(z\right)$ a modified Bessel function of the first
kind, valid for both, large and small $z$ respectively with 
\begin{eqnarray}
\mu\left(z\right) & = & \begin{cases}
-2\ln\left(z/2\right) & \text{\ for }z\ll1\\
z^{-1} & \text{\ for }z\gg1
\end{cases},\label{eq:mu}
\end{eqnarray}
 Eq.(\ref{eq:Action_JJ_2}) can be written as 
\begin{multline}
S\left[\phi,n\right]=\sum_{\tilde{\tau},r,r'}2\Delta\tau n_{0}\left(\tilde{\tau},r\right)\left[C_{R}^{-1}\right]_{rr'}n_{0}\left(\tilde{\tau},r'\right)\\
+\sum_{\tilde{\tau},r}\left\{ \frac{1}{2}\mu\left(\frac{I_{c}\Delta\tau}{2}\right)n_{1}^{2}\left(\tilde{\tau},r\right)-i\phi_{r}^{t}n_{1}\left(\tilde{\tau},r\right)-i\phi\left(\tilde{\tau},r\right)\times\right.\\
\left.\times\left[n_{1}\left(\tilde{\tau},r-1\right)-n_{1}\left(\tilde{\tau},r\right)+n_{0}\left(\tilde{\tau}-1,r\right)-n_{0}\left(\tilde{\tau},r\right)\right]\right\} \label{eq:Action_JJ_3}
\end{multline}
where we relabel $n\to n_{0}$ in Eq.(\ref{eq:Action_JJ_2}) and $m\to n_{1}$
in Eq.(\ref{eq:Villain}) in order to interpret $n_{\mu}\left(\tilde{\tau},r\right)$
as an integer field living on links of a square lattice - an integer-valued
one-form on the square lattice - with $n_{0}$ corresponding to time-like
and $n_{1}$ to space-like links. 

\begin{figure}[H]
\centering{}\includegraphics[width=0.6\columnwidth]{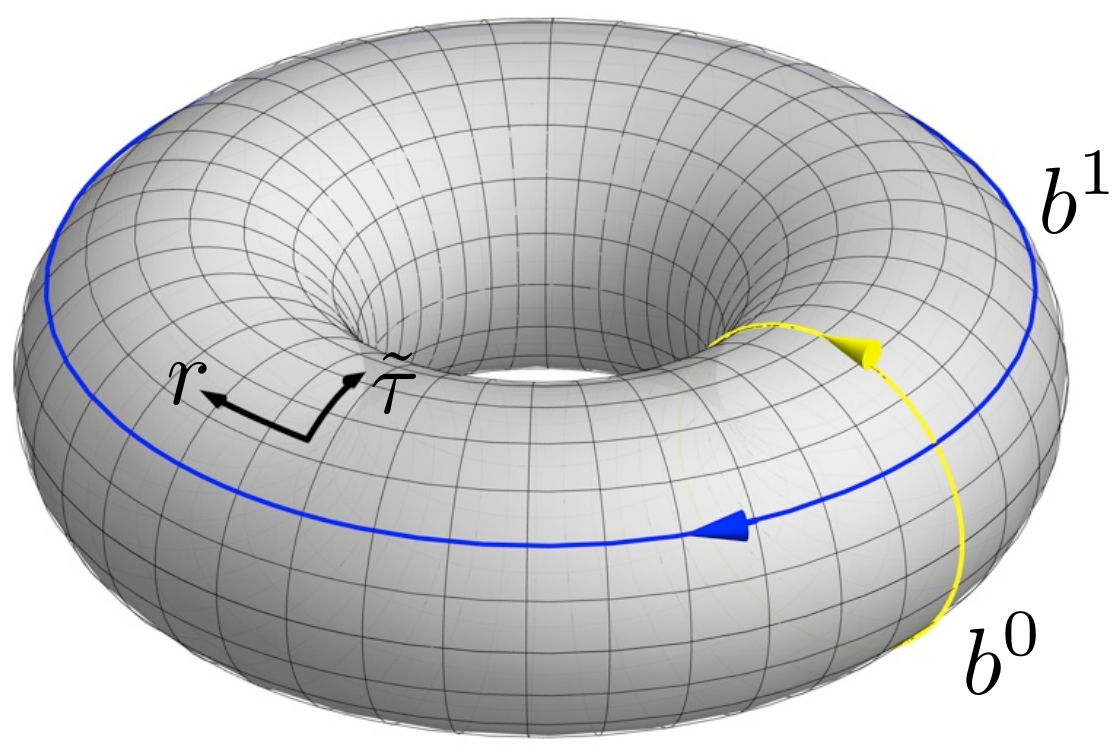}\caption{\label{fig:system-1}Sketch of two integer-valued 1-forms that cannot
be written as $\pd a=\left\{ -\Delta_{1}a_{01},\Delta_{0}a_{01}\right\} $
with $a$ a 2-form. $b_{0}^{0}\left(\tilde{\tau},r\right)=1$ and
$b_{1}^{1}\left(\tilde{\tau},r\right)=1$ for $\left(\tilde{\tau},r\right)$
in the yellow and blue lines respectively otherwise $b_{0}^{0}\left(\tilde{\tau},r\right)=b_{1}^{1}\left(\tilde{\tau},r\right)=b_{0}^{1}\left(\tilde{\tau},r\right)=b_{1}^{0}\left(\tilde{\tau},r\right)=0$
. }
\end{figure}

Integrating out the $\phi$ field yields the divergence-free constraint
\begin{eqnarray}
\pd n & \equiv & \Delta_{1}n_{1}+\Delta_{0}n_{0}=0,\label{eq:del}
\end{eqnarray}
where $\Delta_{0}f\left(\tilde{\tau},r\right)=f\left(\tilde{\tau},r\right)-f\left(\tilde{\tau}-1,r\right)$
is the discrete derivative along the time direction. Locally such
constrain can be satisfied by writing $n$ as the rotational of an
integer valued field living on the centers of plaquettes - an integer-valued
lattice 2-form - $n=\pd a$ or in components: $n_{0}=-\Delta_{1}a_{01}$,
$n_{1}=\Delta_{0}a_{01}$, where the subscript of $a$ denotes that
this field lives on spacial-temporal plaquettes. The operator $\pd$
can be seen as the lattice exterior coderivative. Globally, the most
generic solution of the constraint in Eq.(\ref{eq:del}) includes
a non-trivial divergence-free field that cannot be written as a rotational.
On a torus, such general solution can be decomposed as $n=\pd a+\sum_{\alpha}c_{\alpha}b^{\alpha}$.
More explicitly,
\begin{eqnarray}
n_{0}\left(\tilde{\tau},r\right) & = & -\Delta_{1}a_{01}\left(\tilde{\tau},r\right)+\sum_{\alpha=0,1}c_{\alpha}b_{0}^{\alpha}\left(\tilde{\tau},r\right)\\
n_{1}\left(\tilde{\tau},r\right) & = & \Delta_{0}a_{01}\left(\tilde{\tau},r\right)+\sum_{\alpha=0,1}c_{\alpha}b_{1}^{\alpha}\left(\tilde{\tau},r\right)
\end{eqnarray}
where $b^{0}$ and $b^{1}$ (with $\pd b^{\alpha}=0$) are integer-valued
1-forms on the lattice that cannot be written as a rotational. They
are chosen, see Fig.(\ref{fig:system-1}), to have a minimum flux
along time and space directions respectively: $\sum_{\tilde{\tau},r}b_{\mu}^{0}\left(\tilde{\tau},r\right)=N\delta_{\mu0}$,
$\sum_{\tilde{\tau},r}b_{\mu}^{1}\left(\tilde{\tau},r\right)=L\delta_{\mu1}$.
$c_{\alpha=0,1}$ are integer-valued coefficients labeling different
topological sectors. Note that in the infinite volume limit, i.e.
zero temperature and $L\to\infty$, the $b$ terms can be dropped
in the solution as the space becomes topologically trivial. Later
on we will drop the $b^{0}$ contribution as we are interested in
the zero temperature limit. 

In terms of the $a$ field and the integers $c{}_{0}$ and $c_{1}$,
the partition function is given by the unconstrained sum $Z=\sum_{a,c}e^{-S\left[a,c\right]}$
with
\begin{align}
S\left[a,c\right] & =\sum_{\tilde{\tau},r,r'}2\Delta\tau\left[\Delta_{1}a\left(\tilde{\tau},r\right)-\sum_{\alpha}c_{\alpha}b_{0}^{\alpha}\left(\tilde{\tau},r\right)\right]\times\nonumber \\
 & \left[C_{R}^{-1}\right]_{rr'}\left[\Delta_{1}a\left(\tilde{\tau},r'\right)-\sum_{\alpha}c_{\alpha}b_{0}^{\alpha}\left(\tilde{\tau},r'\right)\right]\nonumber \\
 & -ic_{1}\Phi+\frac{1}{2}\sum_{\tilde{\tau},r}\mu\left(\frac{I_{c}\Delta\tau}{2}\right)\left[\Delta_{0}a\left(\tilde{\tau},r\right)+c_{\alpha}b_{1}^{\alpha}\left(\tilde{\tau},r\right)\right]^{2}\label{eq:Action_JJ_4}
\end{align}
where the total flux $\Phi=\sum_{r}\phi_{r}^{t}$. 

Using the Poisson summation formula $\sum_{a}f\left(a\right)=\sum_{m}\int d\psi\, f\left(\psi\right)e^{2\pi im\psi}$
to improve the convergence of the sum over Eq.(\ref{eq:Action_JJ_4})
\cite{jose} and integrating over $\psi$ yields
\begin{eqnarray}
Z & = & \sum_{m,c}\delta_{\sum m=0}e^{-S\left[c,m\right]}e^{ic_{1}\Phi}\label{eq:Z_JJ_5}
\end{eqnarray}
where the sum over $m$ is restricted such that the so-called neutrality
condition $\sum_{r\tilde{\tau}}m\left(\tilde{\tau},r\right)=0$ is
fulfilled \cite{jose} and 
\begin{eqnarray}
S\left[c,m\right] & = & \frac{\left(2\pi\right)^{2}}{2}\sum_{\tilde{\tau}\tilde{\tau}'rr'}m\left(\tilde{\tau},r\right)G\left(\tilde{\tau}-\tilde{\tau}',r-r'\right)m\left(\tilde{\tau}',r'\right)\nonumber \\
 &  & -2\pi i\sum_{\alpha}c_{\alpha}\sum_{\tilde{\tau},r}m\left(\tilde{\tau},r\right)\left(\pd^{-1}b^{\alpha}\right)\left(\tilde{\tau},r\right).\label{eq:Action_JJ_5}
\end{eqnarray}
with $\left(\pd^{-1}b^{\alpha}\right)_{01}=\left(\Delta_{0}^{2}+\Delta_{1}^{2}\right)^{-1}\left(\Delta_{1}b_{0}^{\alpha}-\Delta_{0}b_{1}^{\alpha}\right)$
the inverse of the $\pd$ operator defined in Eq.(\ref{eq:del}).
The last term in Eq.(\ref{eq:Action_JJ_5}) for $b^{1}$can be simplified
to 
\begin{eqnarray*}
\sum_{\tilde{\tau},r}m\left(\tilde{\tau},r\right)\left(\pd^{-1}b^{1}\right)\left(\tilde{\tau},r\right) & = & \sum_{j}\left[\left(\Delta_{0}^{2}+\Delta_{1}^{2}\right)^{-1}\bar{\Delta}_{0}m\right]\left(0,r\right).
\end{eqnarray*}

The Green's function is given by
\begin{multline}
G^{-1}=-4\Delta\tau\tilde{C}_{\text{S}}^{-1}\Delta_{1}^{2}\left[\frac{1}{\left\{ 1-C_{\text{M}}^{-1}\tilde{C}_{\text{S}}\Delta_{1}^{2}\right\} }-\tilde{C}_{\text{S}}^{-1}C_{\text{J}}\Delta_{1}^{2}\right]^{-1}\\
-\mu\left(\frac{I_{c}\Delta\tau}{2}\right)\Delta_{0}^{2}\label{eq:G_JJ_5}
\end{multline}
In summary, after integrating over the $\psi$ field that represents
small phase fluctuations, the action in Eq.(\ref{eq:Action_JJ_5})
is given solely in terms of topological excitations, $m$, that can
be interpreted as an instanton field representing a phase slip. The
corrections due to non-vanishing values of $C_{\text{S}}^{-1}C_{\text{J}}$
and $C_{\text{M}}^{-1}\tilde{C}_{\text{S}}$ do not change the nature
of the long-range interaction between the phase slips, as they multiply
higher powers of the discrete Laplacian. Nonetheless they appear in
Eq.(\ref{eq:Action_JJ_5}) in inequivalent ways, further we will see
this translates to different contribution to the monopoles energy
to create monopole pairs.

\subsection{Flux quantization}

To understand how the flux piercing the ring gets quantized in the
superconducting phase, where the density of instantons (phase slips)
vanishes, let us examine the partition function given in Eq.(\ref{eq:Z_JJ_5}).
For simplicity let us first take the zero temperature limit in order
to ignore the $b^{0}$ field. The flux $\Phi$ is imposed to the system
assuming that the magnetic field far from the ring is constant and
perpendicular to the $z$ axes in Fig.(\ref{fig:system}). A complete
description of the system array+field should include the dynamics
of the electromagnetic field as well. However this is too involved
and not really needed here, the only thing that is required is to
remember that the spacial distribution of the electromagnetic field
(and thus the flux piercing the ring) is itself determined by an action
containing the electromagnetic contribution plus the coupling of the
electromagnetic field to the instanton configurations given by the
last term of Eq.(\ref{eq:Action_JJ_5}). 

Performing the summation over $c_{1}$ in Eq.(\ref{eq:Z_JJ_5}) one
observes that the partition function of a system with flux $\Phi$
can be written as 
\begin{multline}
Z=\sum_{m,c_{1}}\delta_{\sum m=0}\ \delta_{2\pi}\left(\Phi-\Phi_{m}^{1}\right)\\
\times e^{-\frac{\left(2\pi\right)^{2}}{2}\sum_{\tilde{\tau}r,\tilde{\tau}'r'}m\left(\tilde{\tau},r\right).G\left(\tilde{\tau}-\tilde{\tau}',r-r'\right).m\left(\tilde{\tau}',r'\right)}\label{eq:Z_JJ_6}
\end{multline}
where $\delta_{2\pi}$ is the $2\pi$-periodic delta function and
$\Phi_{m}^{1}=2\pi\sum_{r}\left[\frac{\bar{\Delta}_{0}}{\Delta_{0}^{2}+\Delta_{1}^{2}}m\right]\left(0,r\right)\in\mathbb{R}$.
To the action of the free electromagnetic action one should thus add
the monopole contribution $F\left[\Phi\right]=-\ln Z$. Directly from
Eq.(\ref{eq:Z_JJ_6}) one can observe that if the density of phase-slips
vanishes (i.e. $\av{\frac{1}{NL}\sqrt{\sum_{\tilde{\tau},r}m^{2}\left(\tilde{\tau},r\right)}}=0$)
then $\Phi_{m}^{1}=0$ and thus $\Phi$ has to be quantized in multiples
of $2\pi$. When phase-slips proliferate, $\Phi_{m}^{1}$ is a fraction
of $2\pi$, for a generic configuration of instantons $m$, the summation
over all $m$ configurations allows for a continuum value of $\Phi$.

\subsection{Superconducting-Insulating Transition}

\begin{figure*}
\centering{}\includegraphics[width=0.9\textwidth]{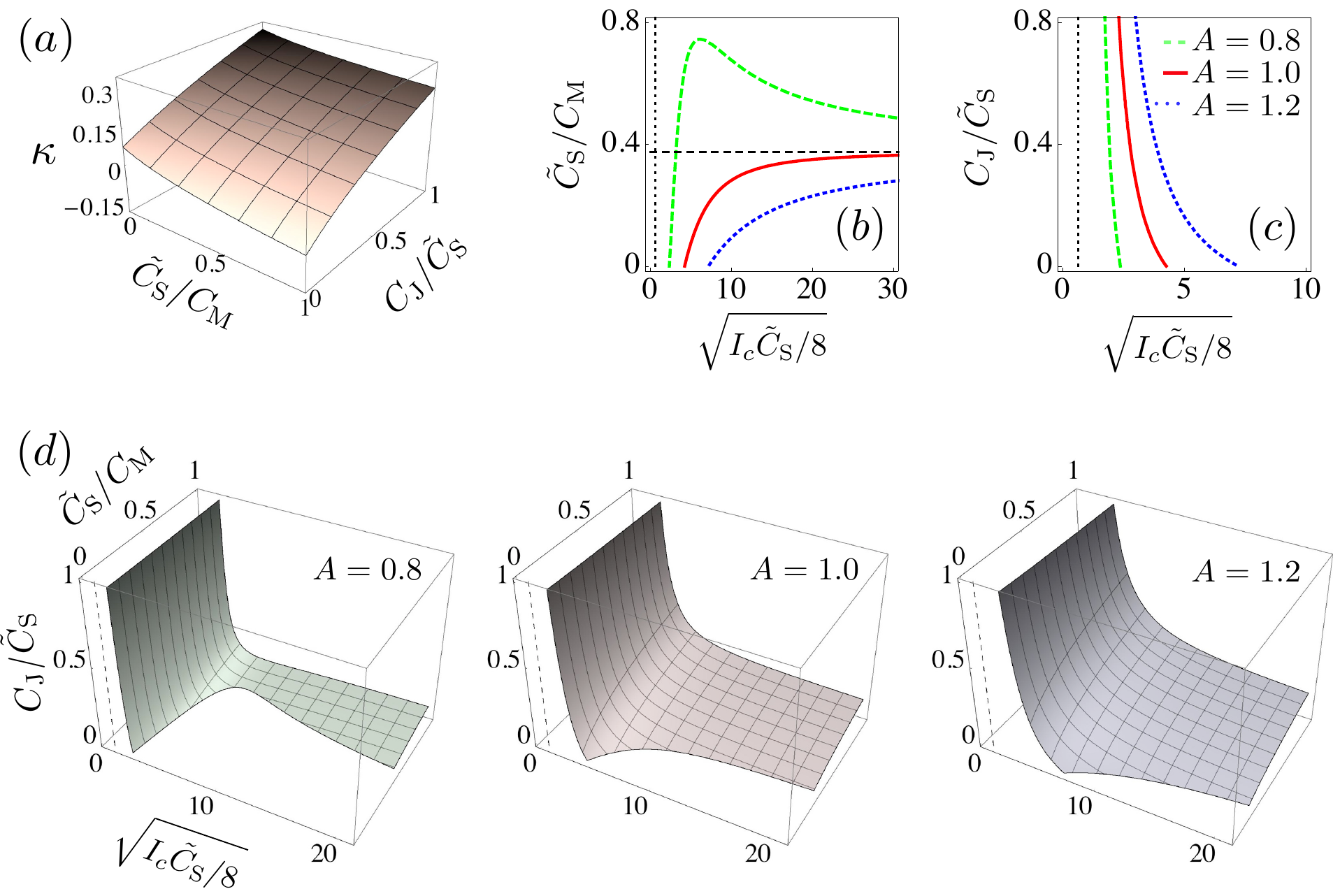}\caption{\label{fig:transition}(a) Plot of the function $\kappa$ as a function
of $\tilde{C}_{\text{S}}/C_{\text{M}}$ and $C_{\text{J}}/\tilde{C}_{\text{S}}$
computed numerically form the asymptotic form of $G\left(\tilde{\tau},r\right)-G\left(0,0\right)$
for $\tau,r\to\infty$ at $\lambda=1$. (b) Phase diagram in the $\{\sqrt{I_{c}\tilde{C}_{\text{S}}/8},\tilde{C}_{\text{S}}/C_{\text{M}}\}$
plane for $C_{\text{J}}=0$ for different values of the non-universal
constant $A$. Below the phase transition line the system is a superconductor
and above it is insulator. The horizontal dashed line corresponds
to the critical ratio $\tilde{C}_{\text{S}}/C_{\text{M}}\approx0.375$
above which the system is always in the insulating phase in the $\sqrt{I_{c}\tilde{C}_{\text{S}}/8}\to\infty$
limit. The vertical dotted line at $\tilde{C}_{\text{S}}/C_{\text{M}}=2/\pi$
marks the lower bound obtained for $A=0$. (c) Phase diagram in the
$\{\sqrt{I_{c}\tilde{C}_{\text{S}}/8},C_{\text{J}}/\tilde{C}_{\text{S}}\}$
plane for $C_{\text{M}}\to\infty$. (d) Complete phase diagram in
the $\{\sqrt{I_{c}\tilde{C}_{\text{S}}/8},\tilde{C}_{\text{S}}/C_{\text{M}},C_{\text{J}}/\tilde{C}_{\text{S}}\}$
space for different values of $A$. }
\end{figure*}

Having understood how the flux gets quantized once instantons are
suppressed, let us neglect the topological terms (i.e. set $c_{0,1}=0$
in Eq.(\ref{eq:Z_JJ_5})), in order to study the superconducting-insulating
transition. A simple way of addressing this question is to map the
problem to the Sine-Gordon model. The main result we report in this
section is that the superconducting insulating phase transition is
Kosterlitz-Thouless like, even in the presence of a finite $C_{\text{M}}$
and $C_{\text{J}}$. This extends the results of Ref.\cite{newcap},
where the case $C_{\text{M}}\neq0$, $C_{\text{J}}=0$ is considered.
Nonetheless $C_{\text{M}}$ and $C_{\text{J}}$ renormalize the instanton-core
energy in rather different ways. By studying how this energy gets
renormalized we obtain the behavior of the superconducting-insulating
transition line as a function of $I_{c}$, $\tilde{C}_{\text{S}}$,
$C_{\text{M}}$ and $C_{\text{J}}$. We note that $\tilde{C}_{\text{S}}$
also includes a term $\propto1/\delta$ coming from quantum fluctuations
induced by finite size effects that so far had not investigated in
the literature. 

The first step to get the Sine-Gordon action is to regularize the
instanton interaction kernel at the origin $G\left(\tilde{\tau},j\right)\to G\left(\tilde{\tau},j\right)-G\left(0,0\right)$
in Eq.(\ref{eq:Action_JJ_5}) by making use of the neutrality condition.
After this procedure the asymptotic $\tilde{\tau},j\to\infty$ form
of the instanton (anti) instanton interaction is given by
\begin{multline}
\left(2\pi\right)^{2}\left[G\left(\tilde{\tau},r\right)-G\left(0,0\right)\right]\simeq\\
\tilde{G}\left(\tilde{\tau}-\tilde{\tau}',r-r'\right)-\nu,\label{eq:nu_def}
\end{multline}
where 
\begin{eqnarray}
\tilde{G}\left(\tilde{\tau},r\right) & = & -2\pi\sqrt{I_{c}\tilde{C}_{\text{S}}/8}\ln\left(\sqrt{\tilde{\tau}^{2}/\lambda^{2}+r^{2}}\right)
\end{eqnarray}
 is the long-range instanton interaction potential and 
\begin{eqnarray}
\nu & = & \sqrt{I_{c}\tilde{C}_{\text{S}}/8}\ \kappa\left(\lambda,\tilde{C}_{\text{S}}/C_{\text{M}},C_{\text{J}}/\tilde{C}_{\text{S}}\right)
\end{eqnarray}
 is the instanton-core energy. Choosing the regulator $\Delta\tau\approx\sqrt{\tilde{C}_{\text{S}}/2I_{c}}$
such that $ $$I_{c}\Delta\tau/2\gg1$ and $\Delta\tau\ll1$ \cite{fazio2,koreans}
, we observe by Eq.(\ref{eq:mu}) that $\mu\left(\frac{I_{c}\Delta\tau}{2}\right)\simeq\frac{2}{I_{c}\Delta\tau}$.
The anisotropy between time and space directions $ $$\lambda=\sqrt{\tilde{C}_{\text{S}}/2I_{c}\left(\Delta\tau\right)^{2}}$
is thus of order 1. $\kappa$ depends on all ratios $\lambda,\tilde{C}_{\text{S}}/C_{\text{M}}$
and $C_{\text{J}}/\tilde{C}_{\text{S}}$, however it is mildly varying
as a function of $\lambda$ around $\lambda=1$. In the following
we take the $\lambda=1$ prescription \cite{fazio2,koreans} for our
numerical analysis. 

The function $\kappa$ can be computed numerically by subtracting
the asymptotic behavior $\tilde{G}\left(\tilde{\tau},r\right)$ to
the right hand site of Eq.(\ref{eq:nu_def}) and numerically integrating
the resulting expression. After a careful analysis of the numerical
data to ensure that the asymptotic values are well reproduced we obtained
the results of Fig.(\ref{fig:transition})-(a). 

Using the neutrality condition once more, the action acquires the
Coulomb (lattice) gas form 
\begin{multline}
S\left[m\right]\simeq\sum_{r\tilde{\tau}\neq r'\tilde{\tau}'}m\left(\tilde{\tau},r\right)\tilde{G}\left(\tilde{\tau}-\tilde{\tau}',r-r'\right)m\left(\tilde{\tau}',r'\right)\\
+\nu\sum_{r\tilde{\tau}}\left[m\left(\tilde{\tau},r\right)\right]^{2}
\end{multline}
The (lattice) Sine-Gordon model can be obtained by inserting an Hubbard-Stratonovich
field and using the identity given in Eq.(\ref{eq:Villain}): 
\begin{eqnarray}
Z & = & \sum_{m}\delta_{\sum m=0}e^{-S\left[m\right]}\nonumber \\
 & \propto & \int D\psi\, e^{-\frac{1}{2}\psi\tilde{G}^{-1}\psi+u\sum_{x}\cos\left(\psi_{x}\right)}
\end{eqnarray}
with $\mu\left(u\right)=\nu$ given by Eq.(\ref{eq:mu}). Note that
in this mapping the neutrality condition is assured by the fact that
$\tilde{G}^{-1}\left(\omega=0,k=0\right)=0$. The usual (continuum)
Sine-Gordon action, that maintains the universal properties of the
lattice model, is obtained taking the continuum limit by formally
introducing a regularizing lattice constant $a$ and taking the limit
$a\to0$. In the continuous form the inverse of the kernel $\tilde{G}$
can be straightforwardly identified: $\frac{1}{2\pi}\left(\frac{1}{\lambda}\pd_{x_{1}}^{2}+\lambda\pd_{x_{0}}^{2}\right)\ln\left(\sqrt{\frac{x_{0}^{2}}{\lambda^{2}}+x_{1}^{2}}\right)=\delta\left(x_{0}\right)\delta\left(x_{1}\right)$.
After a rescaling of the axes in the $x_{0}$ direction, one obtains
the continuum Sine-Gordon action 
\begin{eqnarray}
S & = & -\frac{1}{2}\int d^{2}x\left[g\left(\nabla\psi\right)^{2}-\lambda a^{-2}u\cos\left(\psi\right)\right]
\end{eqnarray}
with $g=\frac{1}{\left(2\pi\right)^{2}}\sqrt{\frac{8}{I_{c}\tilde{C}_{\text{S}}}}$.
This model has a phase transition for $g=g_{c}$, that can be estimated
by a perturbative renormalization group procedure to first order in
$u$ \cite{sinegordon}: 
\begin{eqnarray}
g_{c} & = & \frac{1}{8\pi}-y_{1}\lambda u+O\left(u^{2}\right)\label{eq:GS_phase_transition}
\end{eqnarray}
where $y_{1}\simeq1/8$ \cite{Amit} and $\mu\left(u\right)\simeq-2\ln\left(u/2\right)$. 

Substituting this values in Eq.(\ref{eq:GS_phase_transition}) one
obtains the phase transition condition 

\begin{eqnarray}
\sqrt{\frac{8}{I_{c}\tilde{C}_{\text{S}}}} & = & \frac{\pi}{2}\left[1-Ae^{-\frac{1}{2}\sqrt{\frac{I_{c}\tilde{C}_{\text{S}}}{8}}\kappa\left(\lambda,\frac{\tilde{C}_{\text{S}}}{C_{\text{M}}},\frac{C_{\text{J}}}{\tilde{C}_{\text{S}}}\right)}\right]\label{eq:transition line}
\end{eqnarray}
where $A=16\pi y_{1}\lambda$.

Eq.(\ref{eq:transition line}) predicts the form of the Kosterlitz-Thouless
transition line as a function of $\tilde{C}_{\text{S}}/C_{\text{M}}$,
$C_{\text{J}}/\tilde{C}_{\text{S}}$ and the non-universal constant
$A$. We have now all the ingredients to discuss the phase diagram
of the 1D JJ array.

\section{Discussion}

The phase diagram as a function of $\tilde{C}_{\text{S}}/C_{\text{M}}$
for $C_{\text{J}}=0$ is depicted in Fig.(\ref{fig:transition})-(b).
As was expected the stability of the superconducting phase is reduced
upon increasing the ratio $\tilde{C}_{\text{S}}/C_{\text{M}}$, in
agreement with Ref. \cite{koreans} where a perturbative analysis
around $\tilde{C}_{\text{S}}/C_{\text{M}}=0$ was performed. For $\tilde{C}_{\text{S}}/C_{\text{M}}\to\infty$
it is well known that \cite{doniach} the system is always in the
insulating phase independently of the value of $\sqrt{I_{c}\tilde{C}_{\text{S}}/8}$.
The expression Eq.(\ref{eq:transition line}) interpolates between
this two regimes. It predicts a critical value $\tilde{C}_{\text{S}}/C_{\text{M}}\approx0.375$
above which the system is always in the insulating phase in the $\sqrt{I_{c}\tilde{C}_{\text{S}}/8}\to\infty$
limit. For this critical ratio $\kappa$ vanishes and becomes negative
($\kappa<0$) for larger values of $\tilde{C}_{\text{S}}/C_{\text{M}}$
which, for sufficiently large $\sqrt{I_{c}\tilde{C}_{\text{S}}/8}$,
renders the system insulating due to the proliferation of phase slips.
The non-linearity of the relation Eq.(\ref{eq:transition line}) induces
a striking feature in the transition line for $A$ smaller than unity:
superconductivity is predicted to have a re-entrant behavior. Here,
upon increasing $\sqrt{I_{c}\tilde{C}_{\text{S}}/8}$, the system
passes from insulator to superconductor and again to insulator. This
is a rather contra-intuitive behavior as one would naively expect
that an increase of the Josephson energy (proportional to $I_{c}$)
always enhances superconductivity. It would be very interesting to
search for experimental signatures of this phenomena. However we must
also note that $A$ is a non-universal constant that depends on various
factors including the accuracy to which the instanton fugacity is
computed, the exact choice of $\Delta\tau$ and the system parameters.
At present we cannot rule out that in the range of plausible parameters
for realistic materials $A\geq1$ and this non-monotonicity is not
observed. Another potential limitation of our results is that, since
Eq.(\ref{eq:GS_phase_transition}) is only valid for small values
of $u$, the obtained transition lines are only qualitatively correct. 

As is observed in Fig.(\ref{fig:transition})-(c), the presence of
a finite $C_{\text{J}}$, in the limit $C_{\text{M}}\to\infty$, increases
the stability of the superconducting phase. Even away from this limit,
a finite $C_{\text{J}}$ makes more robust the superconducting phase.
In Fig.(\ref{fig:transition})-(d) it is depicted the full phase diagram
as a function of $\sqrt{I_{c}\tilde{C}_{\text{S}}/8}$, $\tilde{C}_{\text{S}}/C_{\text{M}}$
and $C_{\text{J}}/\tilde{C}_{\text{S}}$ for different values of the
non-universal constant $A$. Another striking feature of the phase
diagram, besides the re-entrant behavior mentioned previously, is
the fact that, even for a relatively large ratio $\tilde{C}_{\text{S}}/C_{\text{M}}$
which brings the system deep into the insulating phase, a fairly small
value of $C_{\text{J}}/\tilde{C}_{\text{S}}$ can restore superconductivity. 

There are also intriguing features related to the interplay between
quantum capacitance and charging effects. For instance in the limit
in which the charging energy is only due to a finite mutual capacitance
there is no global superconductivity \cite{doniach} as phase fluctuations
in each grain are independent. However the inclusion of ``quantum''
capacitance $C_{\delta}$, induced by quantum size effects not related
to Coulomb interactions, changes this picture qualitatively. From
Eq.(\ref{eq:transition line}) it is clear that a finite $C_{\delta}$
might stabilize superconductivity in a certain range of parameters
even if the self-capacitance energy is zero. Therefore a finite ``quantum''
capacitance, which occur in all systems no matter the nature of the
interactions, can help restore long-range order in some cases.

\section{Application to cold atom physics}

\begin{figure}[H]
\centering{}\includegraphics[width=0.9\columnwidth]{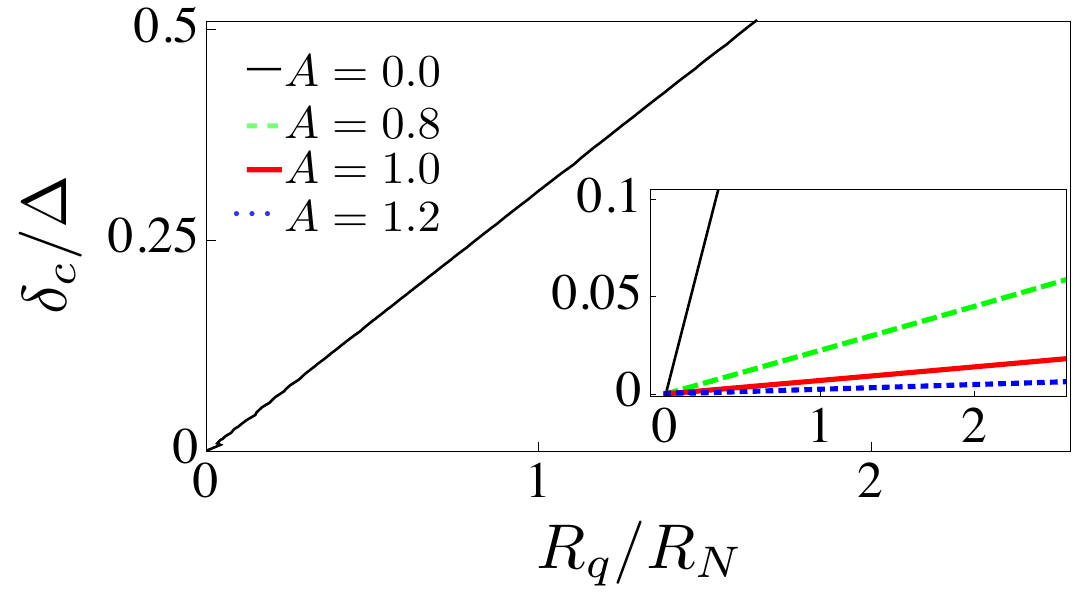}\caption{\label{fig:system-1-1}Phase diagram as a function of $\delta_{c}/\Delta$
and $R_{q}/R_{N}$ plotted for different values of the non-universal
constant $A$. Below the curves the system is superconducting and
above it behaves as an insulator. }
\end{figure}

In this section we investigate the fate of superconductivity in an
array in which Coulomb interactions are absent in the limit in which
the grain mean level spacing $\delta$ becomes comparable to the bulk
gap. For that purpose we study the interplay between the Josephson
coupling, the quantum capacitance $C_{\delta}\sim2/\delta$, and the
quasiparticle dissipation $C_{J}$. This question can be easily addressed
by solving Eq.(\ref{eq:transition line}) in the limit of negligible
charging energies. This is not of academic interest as it is possible
to study experimentally 1D JJ arrays in a cold atom setting \cite{cold}
with no Coulomb interactions at all. Moreover in cold atom physics
many parameters such the tunneling rate, directly related to $R_{N},$
and the gap $\Delta_{0}$ can be controlled with great precision so
an experimental verification seems feasible. 

For sufficiently small grains it is broadly expected that superconductivity
will not survive unless the grains are strongly coupled so that the
effective granularity of the array is heavily suppressed. Likewise
we expect to have global superconductivity for large grains where
quantum fluctuations are negligible. Therefore for a given value of
the normal resistance $R_{N}$ there must exist a minimum grain size
for which phase coherence can occur despite a finite ``quantum''
capacitance $C_{\delta}$. According to Eq.(\ref{eq:transition line}),
the best case scenario for the array to stay superconducting corresponds
to the limit of infinite fugacity (or $A=0$) which sets the following
lower bound on the grain mean level spacing $\delta_{c}\approx\frac{\pi I_{c}}{32}=\frac{\pi^{2}\Delta_{0}R_{q}}{32R_{N}}$
from which it is possible to estimate the minimum grain size. For
metallic superconductors the above estimation results in a minimum
grains size is of order $L\sim5$nm though important variations are
expected depending on the material. A finite fugacity is expected
to weaken the superfluid state and therefore to decrease $\delta_{c}$.
The evolution of $\delta_{c}$ as a function of $R_{q}/R_{N}$ for
different values of $C_{J}$ and the non-universal parameter $A$,
depicted in Fig. 4, agree with this prediction. Note that no re-entrant
behavior is observed as there is no charging energy related to a mutual
capacitance . Finally we note that our calculation is only valid for
$\delta/\Delta_{0}\ll1$ so, from the above expression for $\delta_{c}$,
it is clear that phase coherence is attainable even in the region
$R_{N}\sim R_{q}$ where the contact among grains is weak and only
induces a small smoothing of the spectral density.

\section{Size dependence of classical and ``quantum'' capacitance}

As the grain size decreases both classical and quantum capacitance
play a more important role in the description of the array. Naively
one might think that for sufficiently small grains charging effects
are in general less important than quantum capacitance effects since
the former $E_{c}\propto1/L^{2}$ but the latter is proportional to
$\delta\propto1/L^{3}$. However we note the capacitance and the mean
level spacing depends on completely different parameters, the former
on the dielectric constant of the material and the details of the
substrate while the latter on the Fermi energy and the effective electronic
mass. As a result it is plausible that, even if the area scaling holds,
both contributions might still be similar for grain sizes $L\sim10$nm.
This is consistent with the experimental results of \cite{wolf} for
Pb superconducting islands where it was possible to reproduce the
expected classical scaling of the capacitance with the area only for
relatively large grains. Indeed in a Si(111) substrate the charging
energy and the mean level spacing of a $L\sim7$nm grain with $C\approx40$aF
can be comparable. Therefore quantum fluctuations, not related to
charging effects, must be taken into account in any quantitative theoretical
model of superconducting nanograins.

\section{Conclusions}

We have investigated the robustness of superconductivity in a 1D JJ
array of nanograins at zero temperature. We go beyond the standard
theoretical treatment of this problem by including quantum fluctuations,
not related to Coulomb interactions, induced by finite size effects,
referred to as ``quantum capacitance''. By using path integral techniques
we have studied the phase diagram of this system including also charging
effects and quasiparticle dissipation. We have treated the model analytically
by mapping it onto a 1+1D Coulomb gas and then to a sine Gordon model
which is known to undergo a Kosterlitz-Thouless transition. For sufficiently
large grains long range order is always robust to small self-capacitance
charging effects. However the combined effect of a vanishing self-capacitance
energy and a finite mutual capacitance energy leads to breaking of
phase coherence. We have shown that even in this limit superconductivity
is stabilized by a quantum capacitance. In systems with vanishing
charging effects, relevant in cold atom experiments, we have shown
that long range order persists up to normal resistances comparable
to the quantum one. We have also identified the minimum grain size
for global superconductivity to occur in this limit. We have found
that the phase diagram resulting from the renormalization group analysis
is to some extent sensitive to specific details of the model embodied
in a non-universal prefactor of the fugacity. As an example, for certain
capacitance configurations, small changes in the pre-factor of the
fugacity can lead to rather contraintuitive results such as a transition
from superconductor to insulator by increasing the Josephson coupling.
\begin{acknowledgments}
AMG acknowledges financial support from PTDC/FIS/111348/2009, a Marie
Curie International Reintegration Grant PIRG07-GA-2010-26817 and EPSRC
grant EP/I004637/1. \end{acknowledgments}

\end{document}